\documentclass{article}
\usepackage{spconf,amsmath,graphicx,hyperref, booktabs, multirow, multicol, amsfonts, float}

\title{Brainprint-Modulated Target Speaker Extraction}
\name{Qiushi Han$^{1, \dagger}$ \qquad Yuan Liao$^{2, \dagger}$ \qquad Youhao Si$^{1}$ \qquad Liya Huang$^{1, \star}$\thanks{$\dagger$ These authors contributed equally to this work}\thanks{$\star$ Corresponding author}}

\address{
    $^{1}$College of Electronic and Optical Engineering \& College of Flexible Electronics (Future Technology), \\ Nanjing University of Posts and Telecommunications, Nanjing, China \\
    $^{2}$School of Artificial Intelligence, School of Data Science, SRIBD, \\ The Chinese University of Hong Kong, Shenzhen, China
    }
\begin{document}
\ninept
\maketitle
\begin{abstract}
Achieving robust and personalized performance in neuro-steered Target Speaker Extraction (TSE) remains a significant challenge for next-generation hearing aids. This is primarily due to two factors: the inherent non-stationarity of EEG signals across sessions, and the high inter-subject variability that limits the efficacy of generalized models. To address these issues, we propose Brainprint-Modulated Target Speaker Extraction (BM-TSE), a novel framework for personalized and high-fidelity extraction. BM-TSE first employs a spatio-temporal EEG encoder with an Adaptive Spectral Gain (ASG) module to extract stable features resilient to non-stationarity. The core of our framework is a personalized modulation mechanism, where a unified brainmap embedding is learned under the joint supervision of subject identification (SID) and auditory attention decoding (AAD) tasks. This learned brainmap, encoding both static user traits and dynamic attentional states, actively refines the audio separation process, dynamically tailoring the output to each user. Evaluations on the public KUL and Cocktail Party datasets demonstrate that BM-TSE achieves state-of-the-art performance, significantly outperforming existing methods. Our code is publicly accessible at: https://github.com/rosshan-orz/BM-TSE.
\end{abstract}

\section{Introduction}
\label{sec:intro}

The human brain exhibits a remarkable capacity for selective auditory attention, allowing individuals to focus on a single speaker in complex acoustic environments like a cocktail party\cite{o2015attentional}. Replicating this function is crucial for developing advanced hearing aids for those with hearing loss\cite{hjortkjaer2025real}. Neuro-steered Target Speaker Extraction (TSE) has emerged as a groundbreaking paradigm to address this, aiming to decode a listener's attentional focus directly from their electroencephalography (EEG) signals\cite{ni2024dbpnet, faghihi2022neuroscience,yan2024darnet}.

Significant progress has been made with end-to-end TSE models. Early works like BISS estimated speech envelopes from EEG to guide extraction\cite{ceolini2020brain}, while later models such as U-BESD and BASEN introduced direct time-domain fusion using feature-wise modulation and cross-attention\cite{hosseini2022end,zhang2023basen}. More recent approaches like NeuroHeed(+)\cite{pan2024neuroheed+} and MSFNet\cite{fan2024msfnet} have further improved performance by incorporating attention backends and graph neural networks. However, these methods face a key limitation in their feature encoding. They often process either temporal or spatial characteristics of the EEG in isolation, while a robust fusion of spatio-temporal features is needed to more effectively capture the brain's dynamic response to auditory stimuli. Furthermore, the inherent non-stationarity of EEG signals remains a major challenge, limiting model reliability across different sessions.

Beyond feature encoding, a more profound conceptual gap exists in current TSE models. It is well-established that EEG signals contain unique and stable person-specific patterns, giving rise to the field of brainprint recognition\cite{saha2020intra,yang2022person,wang2020brainprint}. However, most TSE frameworks are designed to learn a generalized mapping from EEG to speech. These models implicitly treat this rich, identity-specific neural information as statistical variance to be suppressed, rather than a valuable signal to be exploited.

To address these challenges, we propose the Brainprint-Modulated Target Speaker Extraction (BM-TSE) framework, a novel architecture designed for robust, personalized, and high-fidelity speech extraction. The foundation of our model is a robust spatio-temporal EEG encoder featuring an Adaptive Spectral Gain (ASG) module. This module is specifically engineered to extract stable and discriminative features from non-stationary signals. Building upon this representation, we introduce a Personalized Brainprint Module that learns a unified brainmap embedding, which is supervised to jointly encode the user's static identity (via SID) and dynamic attentional state (via AAD). The central innovation of our framework is a brainprint modulation mechanism, where this learned brainmap is leveraged to actively refine the separated audio features, dynamically tailoring the output to the specific user.

The main contributions of this work are summarized as follows:
\begin{itemize}
    \item We design a robust, ASG-enhanced spatio-temporal EEG encoder that explicitly addresses the challenge of signal non-stationarity to extract stable neural features.
    \item We propose a unified brainmap representation, a novel embedding supervised to jointly encode a user's unique identity and their dynamic attentional state.
    \item We conduct extensive experiments demonstrating that our proposed model achieves new state-of-the-art results on cocktail party and KUL dataset, significantly outperforming existing TSE methods.
\end{itemize}

\section{Methods}
\label{sec:methods}

\begin{figure*}[ht!]
    \centering
    \includegraphics[width=1\textwidth]{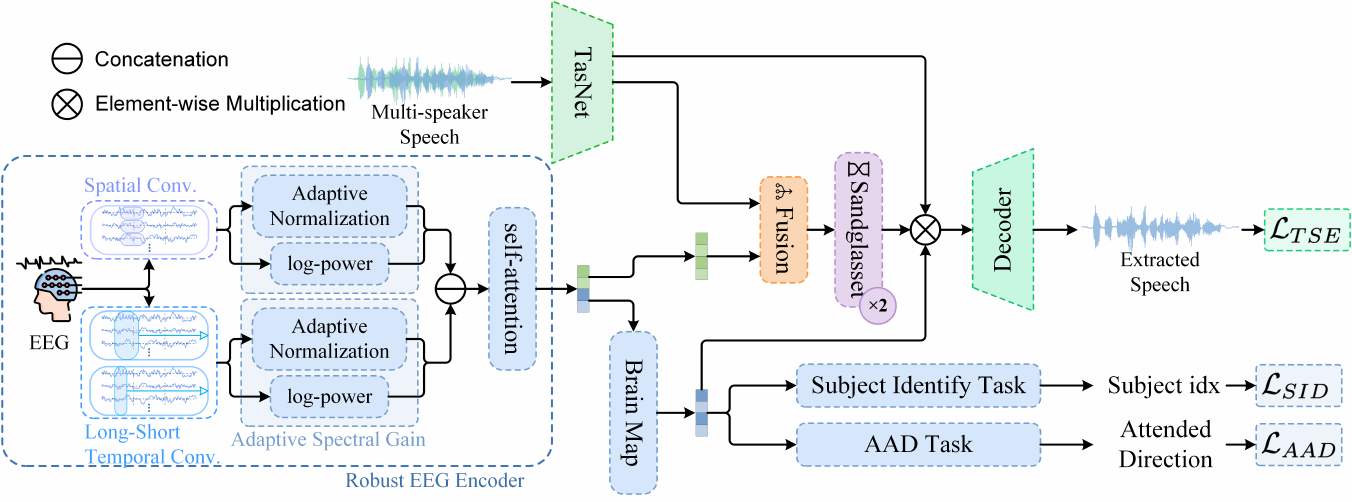}
    \caption{Overview of the BM-TSE architecture. The architecture is an end-to-end system that integrates brain and audio signals. First, a Robust EEG Encoder processes raw EEG to extract a stable feature embedding. This embedding is then fed into a fusion layer alongside multi-speaker audio features, which are processed by a TasNet encoder. The combined features are separated by Sandglasset blocks. The Personalized Brainprint Module creates a brainmap embedding, which refines the separated audio features before the final waveform is rebuilt. The entire model is trained with a multi-task loss function, optimizing for Target Speaker Extraction (TSE), Subject Identification (SID), and Auditory Attention Decoding (AAD) to ensure the brainmap is a personalized and effective guide.}
    \label{fig:architecture}
\end{figure*}

We propose the Brainprint-Modulated Target Speaker Extraction (BM-TSE) framework, an end-to-end architecture for hyper-personalized and high-fidelity speech extraction. As shown in Fig.~\ref{fig:architecture}, the model synergistically optimizes three tasks: target speaker extraction (TSE), subject identification (SID), and auditory attention decoding (AAD). The core idea of BM-TSE is to use a personalized brainmap embedding to directly guide and refine the audio separation process. The framework consists of a robust EEG encoder, a brainprint-guided speaker extractor, and is optimized through a multi-task loss function.

\subsection{Robust EEG Encoder}
\label{ssec:eeg_encoder}

To address the inherent non-stationarity of EEG signals, our encoder is designed to extract robust features suitable for cross-session applications. The encoder processes the input EEG signal $X \in \mathbb{R}^{B \times C \times T}$ by first capturing initial spatio-temporal features, then enhancing them with an Adaptive Spectral Gain (ASG) block, and finally fusing them to produce a unified representation. Here, $B, C, T$ denote the batch size, channel count, and number of time points, respectively.

First, two parallel branches process the input. The Long-Short Temporal Convolution (LS-TConv) branch extracts temporal dynamics, yielding features $E_{\text{temp}} \in \mathbb{R}^{B \times D \times T'}$. Concurrently, the Spatial Convolution (SConv) branch extracts topographical features, producing $E_{\text{spat}} \in \mathbb{R}^{B \times C \times D'}$. Here, $D$ and $D'$ are feature dimensions and $T'$ is the downsampled temporal length.

These initial features are then independently enhanced by the ASG block. For a given input $E_{\text{in}}$, the ASG block contains two sub-components. A log-power block, $L(E_{\text{in}})$, captures nonlinear energy patterns to amplify stable inter-individual differences.
\begin{equation}
    L(E_{\text{in}}) = \log(\text{Pool}(E_{\text{in}}^2 + \epsilon))
\end{equation}
An adaptive normalization block, $A(E_{\text{in}})$, stabilizes feature distributions using a learnable gating mechanism based on Group Normalization (GN).
\begin{equation}
    A(E_{\text{in}}) = E_{\text{in}} \odot \sigma(W_s \odot \text{GN}(E_{\text{in}}) + b_s)
\end{equation}
The outputs are concatenated to form the enhanced representation: $\text{ASG}(E_{\text{in}}) = \text{Concat}(A(E_{\text{in}}), L(E_{\text{in}}))$.

Finally, a cross-domain fusion mechanism integrates the enhanced features. The two feature sets are treated as distinct tokens, enriched with separate learnable positional embeddings ($P_{\text{temp}}, P_{\text{spat}}$), and then concatenated along the sequence dimension. A self-attention layer processes this combined sequence to learn dependencies within and across the domains, producing the final robust EEG feature embedding $E$.
\begin{gather}
\begin{split}
    E'_{\text{temp}} &= \text{ASG}(E_{\text{temp}}) + P_{\text{temp}} \\
    E'_{\text{spat}} &= \text{ASG}(E_{\text{spat}}) + P_{\text{spat}}
\end{split} \\
    E = \text{SelfAttention}(\text{Concat}(E'_{\text{temp}}, E'_{\text{spat}}))
\end{gather}
This final embedding $E$ serves as the input for all subsequent downstream modules.

\subsection{Brainprint-Guided Speaker Extraction}
\label{ssec:extraction}

The final EEG embedding $E$ is used to guide the audio extraction process. The audio pipeline consists of an audio encoder, a separation network, a personalized modulation step, and a final rebuilder.

\textbf{Audio Encoding and Separation.} The input mixed-speech waveform $x$ is first converted into a feature representation $\hat{X}_{\text{audio}}$ by a TasNet encoder~\cite{luo2019conv}. This audio feature is then concatenated with the aligned EEG feature $E$ and fed into a separation network composed of two Sandglasset blocks~\cite{lam2021sandglassetlightmultigranularityselfattentive}.Unlike leading models such as DPRNN\cite{luo2020dual} or DPTNet\cite{zhao2022dptnet}, which operate on a fixed feature granularity, Sandglasset introduces a multi-granularity, sandglass-shaped structure. This design enables the hierarchical modeling of different temporal contexts, such as phonemes and words, which is crucial for effective speech separation.  This network processes the multimodal features and outputs an intermediate separated feature representation, denoted as $A$.

\textbf{Personalized Brainprint Modulation.} Concurrently, the EEG feature $E$ is processed by a lightweight Personalized Brainprint Module consisting of residual convolutional blocks to produce the brainmap embedding. The brainmap is supervised by the SID and AAD losses, forcing it to encode both user-specific traits and attention-related states. The core innovation of our framework is the personalized modulation step, where the brainmap refines the intermediate audio feature $A$:
\begin{equation}
    A_{\text{refined}} = (\mathcal{T}(E) + \mathcal{P}(\text{brainmap})) \odot A
\end{equation}
Here, $\mathcal{T}(\cdot)$ and $\mathcal{P}(\cdot)$ are projection layers that linearly upsample the features to a uniform temporal length. This step allows the brainmap to act as a personalized filter, refining the separation based on the user's unique neural patterns. The refined feature $A_{\text{refined}}$ is then fed to a rebuilder to reconstruct the final time-domain waveform.

\subsection{Multi-Task Optimization}
\label{ssec:loss}

The entire BM-TSE model is trained end-to-end by minimizing a composite multi-task loss function $\mathcal{L}_{\text{total}}$. This loss integrates the objectives from the primary speech extraction task and the two auxiliary tasks that supervise the brainmap embedding. The total loss is defined as:
\begin{equation}
    \mathcal{L}_{\text{total}} = \mathcal{L}_{\text{TSE}} + \alpha \mathcal{L}_{\text{SID}} + \beta \mathcal{L}_{\text{AAD}}
\end{equation}
where $\alpha$ and $\beta$ are hyperparameters that balance the contribution of each task.

\textbf{High-Fidelity TSE Loss ($\mathcal{L}_{\text{TSE}}$).} 
To ensure the extracted waveform $\hat{s}$ has high perceptual quality compared to the ground truth $s$, we use a multi-component loss. $\mathcal{L}_{\text{TSE}}$ is a weighted sum of three metrics that target different aspects of audio fidelity:
\begin{equation}
    \mathcal{L}_{\text{TSE}} = w_1 \mathcal{L}_{\text{MSE}} + w_2 \mathcal{L}_{\text{STFT}} + w_3 \mathcal{L}_{\text{SI-SDR}}
\end{equation}
The components are the time-domain Mean Squared Error ($\mathcal{L}_{\text{MSE}}$), a frequency-domain STFT magnitude loss ($\mathcal{L}_{\text{STFT}}$), and the scale-invariant SI-SDR loss($\mathcal{L}_{\text{SI-SDR}}$).

\textbf{Brainprint Supervision Losses ($\mathcal{L}_{\text{SID}}, \mathcal{L}_{\text{AAD}}$).}
The brainmap embedding is supervised by two auxiliary classification losses. The Subject Identification Loss ($\mathcal{L}_{\text{SID}}$) trains the model to recognize the user's identity. The Auditory Attention Decoding Loss ($\mathcal{L}_{\text{AAD}}$) trains the model to identify the user's current attentional focus. Both tasks are optimized using the standard Cross-Entropy (CE) loss:
\begin{equation}
    \mathcal{L}_{\text{CE}} = - \sum_{c=1}^{C} y_c \log(p_c)
\end{equation}
where $y_c$ is the one-hot ground truth label for class $c$, and $p_c$ is the predicted probability.

\section{Experiments}
\label{sec:experiments}

This section details the experimental setup, including the datasets used, evaluation metrics, and implementation specifics.

\subsection{Datasets}

Our evaluation was conducted on two public multi-speaker EEG-audio datasets: KUL and Cocktail Party.

The \textbf{KUL} dataset \cite{biesmans2016auditory} recorded brain activity from 16 healthy participants as they listened to one of four Dutch short stories. During the experiment, participants were instructed to focus on a specific narrative from two streams presented simultaneously, one to their left and one to their right. The EEG data was captured using a 64-electrode setup at a high sampling rate of 8196 Hz.

The \textbf{Cocktail Party} dataset \cite{broderick2018electrophysiological} included 33 participants who were all healthy and had normal hearing. This study also involved a cocktail party paradigm where two different stories were presented concurrently, with participants instructed to attend to the story in either their left or right ear. Each participant completed 30 trials, each lasting one minute.

\subsection{Implementation Details}
\label{sec:implementation}

Both datasets underwent a similar preprocessing pipeline. The data was band-pass filtered between 0.1 and 45 Hz to isolate relevant neural activity. For the KUL dataset, a notch filter was also used to remove power-line interference, and the data was downsampled to 128 Hz. In the Cocktail Party dataset, channels with excessive noise were corrected using spline interpolation. Re-referencing was performed on both datasets, with the KUL data re-referenced to the average of all channels and the Cocktail Party data to the mastoid channels. Finally, Independent Component Analysis (ICA) \cite{delorme2004eeglab} was applied to remove artifacts from eye movements and muscle activity, ensuring the integrity of the data.

To ensure a fair assessment and evaluate our model's performance, we used a randomized split to partition the data, ensuring samples from both datasets were mixed and then allocated to training, validation, and testing sets. This approach prevents data leakage and provides a more robust evaluation. We used a 75:12.5:12.5 ratio: 75\% of the data was used for training, 12.5\% for validation (where we tracked metrics to save the best-performing model), and the remaining 12.5\% was used as a final, unseen test set.

BM-TSE is implemented using PyTorch and trained on an NVIDIA 4090. We employ the Adam optimizer for model training with an initial learning rate of 1e-4. A `StepLR` scheduler is used to adjust the learning rate, decaying by a factor of 0.9 every epoch. Training is conducted for 100 epochs with a batch size of 8. Training checkpoints are saved periodically, and the model yielding the best SI-SDRi on the validation set is saved as the final model. For more details, refer to the GitHub repository.


To assess our proposed method's effectiveness, we used four key metrics: SI-SDR (dB), PESQ \cite{rix2001perceptual}, STOI \cite{taal2010short}, and ESTOI \cite{jensen2016algorithm}.

\section{Experimental Results}
\label{sec:results}

\subsection{Comparison with Baseline}

We validated the performance of the BM-TSE framework and compared it with several baseline methods. Table~\ref{tab:comparison_cocktail} and~\ref{tab:comparison_kul} presents a comprehensive overview of the performance metrics of all models on the Cocktail Party and KUL.

\begin{table}[htb!]
  \caption{The performance of NeuroSecure TSE compared against competitive baseline models on the Cocktail Party dataset. The evaluation used SI-SDR (dB) to measure speech quality, and STOI, ESTOI, and PESQ to assess speech intelligibility.}
  \label{tab:comparison_cocktail}
  \begin{tabular}{lcccc}
    \toprule
    Model              & SI-SDR (dB) & STOI    & ESTOI   & PESQ    \\
    \midrule
    Mixture           &  0.45       & 0.71    & 0.55    & 1.61    \\
    \midrule
    {UBESD~\cite{hosseini2022end}}     &  \multirow{2}{*}{8.54}       & \multirow{2}{*}{0.83}    & \multirow{2}{*}{--}      & \multirow{2}{*}{1.97}    \\
    \footnotesize TASLP 2022 & & & & \\
    \midrule
    {BASEN~\cite{zhang2023basen}}    & \multirow{2}{*}{11.56}       & \multirow{2}{*}{0.86}    & \multirow{2}{*}{0.72}    & \multirow{2}{*}{2.21}    \\
    \footnotesize Interspeech 2023 & & & & \\
    \midrule
    {MSFNet~\cite{fan2024msfnet}}   & \multirow{2}{*}{12.89}       & \multirow{2}{*}{0.88}    & \multirow{2}{*}{0.77}    & \multirow{2}{*}{\textbf{2.51}}    \\
    \footnotesize ACM MM 2024 &&&& \\
    \midrule
    \textbf{BM-TSE}   & \multirow{2}{*}{\textbf{14.02}}       &   \multirow{2}{*}{\textbf{0.90}}    & \multirow{2}{*}{\textbf{0.77}}    &  \multirow{2}{*}{2.47} \\
    \textbf{(Ours)} &&&& \\
    \bottomrule
  \end{tabular}
\end{table}

\begin{table}[htb!]
  \caption{The performance of NeuroSecure TSE compared against competitive baseline models on the KUL dataset. The evaluation used SI-SDR (dB) to measure speech quality, and STOI, ESTOI, and PESQ to assess speech intelligibility.}
  \label{tab:comparison_kul}
  \begin{tabular}{lcccc}
    \toprule
    Model              & SI-SDR (dB) & STOI    & ESTOI   & PESQ    \\
    \midrule
    Mixture           &  0.25       & 0.69    & 0.52    & 1.17    \\
    \midrule
    {UBESD~\cite{hosseini2022end}}    &  \multirow{2}{*}{6.1}        & \multirow{2}{*}{0.73}    & \multirow{2}{*}{0.75}    & \multirow{2}{*}{1.09}    \\
    \footnotesize TASLP 2022 &&&& \\
    \midrule
    {BASEN~\cite{zhang2023basen}}    & \multirow{2}{*}{11.5}        & \multirow{2}{*}{0.82}    & \multirow{2}{*}{0.76}    & \multirow{2}{*}{1.76}    \\
    \footnotesize Interspeech 2023 &&&& \\
    \midrule
    {MSFNet~\cite{fan2024msfnet}}   & \multirow{2}{*}{14.6}        & \multirow{2}{*}{0.83}    & \multirow{2}{*}{0.76}    & \multirow{2}{*}{\textbf{2.12}}    \\
    \footnotesize ACM MM 2024 &&&& \\
    \midrule
    \textbf{BM-TSE}   & \multirow{2}{*}{\textbf{15.92}}       & \multirow{2}{*}{\textbf{0.85}}    & \multirow{2}{*}{\textbf{0.77}}    & \multirow{2}{*}{2.10}  \\
    \textbf{(Ours)} &&&& \\
    \bottomrule
  \end{tabular}
  \vspace{-1pt}
\end{table}

As shown in the Tables, the proposed BM-TSE model consistently outperforms all other baseline models on both datasets in terms of speech quality (measured by SI-SDR) and speech intelligibility (measured by STOI and ESTOI). Specifically, BM-TSE achieves the highest SI-SDR, STOI, and ESTOI scores, with SI-SDR values of 14.02 dB on the Cocktail Party dataset and 15.92 dB on the KUL dataset. While the MSFNet model shows slightly higher PESQ scores, the overall results demonstrate the superior performance of BM-TSE in improving both speech quality and intelligibility.


\subsection{Ablation Study}

The ablation study in Table \ref{tab:ablation} provides a clear analysis of our proposed BM-TSE model. The complete model achieves top performance, setting a strong benchmark. The results reveal that the LS-TConv module is indispensable, as its removal causes a dramatic performance collapse, with SI-SDRi, a score measuring how much a separation algorithm improves the separation of a target signal compared to the original mixture, falling from 14.50 dB to just 2.88 dB. While less severe, the absence of the ASG and SConv modules also leads to a notable performance decline, underscoring their importance. Additionally, the auxiliary loss function, $\mathcal{L}_{SID}$, proves to be a crucial element, as its removal results in a clear performance degradation.

\begin{table}[htb!]
  \caption{Ablation Study Analysis of Key Modules in the BM-TSE.}
  \label{tab:ablation}
  \centering
  \begin{tabular}{lcccc}
    \toprule
    Model              & SI-SDRi (dB) & STOI    & ESTOI   & PESQ    \\
    \midrule
    BM-TSE    &   \multirow{2}{*}{14.50}  &   \multirow{2}{*}{0.90}  &   \multirow{2}{*}{0.77}  &   \multirow{2}{*}{2.47}  \\
    (Full)&&&& \\
    \midrule
    BM-TSE  &    \multirow{2}{*}{2.88}    &   \multirow{2}{*}{0.72}    &   \multirow{2}{*}{0.54}    &   \multirow{2}{*}{1.70}  \\
    (w/o LS-TConv) &&&& \\
    \midrule
    BM-TSE  &   \multirow{2}{*}{13.61}   &   \multirow{2}{*}{0.88}    &   \multirow{2}{*}{0.74}    &   \multirow{2}{*}{2.37}  \\
    (w/o SConv) &&&& \\
    \midrule
        BM-TSE &   \multirow{2}{*}{13.13}  &   \multirow{2}{*}{0.88}  &   \multirow{2}{*}{0.74}  &   \multirow{2}{*}{2.39}  \\
        (w/o ASG)&&&& \\
    \midrule
    BM-TSE & \multirow{2}{*}{12.29}   &   \multirow{2}{*}{0.87}    &   \multirow{2}{*}{0.72}    &   \multirow{2}{*}{2.26}  \\
    (w/o $\mathcal{L}_{SID}$)&&&& \\
    \bottomrule
  \end{tabular}
\end{table}

\subsection{Visualization}

\begin{figure}[ht]
    \centering
    \includegraphics[width=0.95\linewidth]{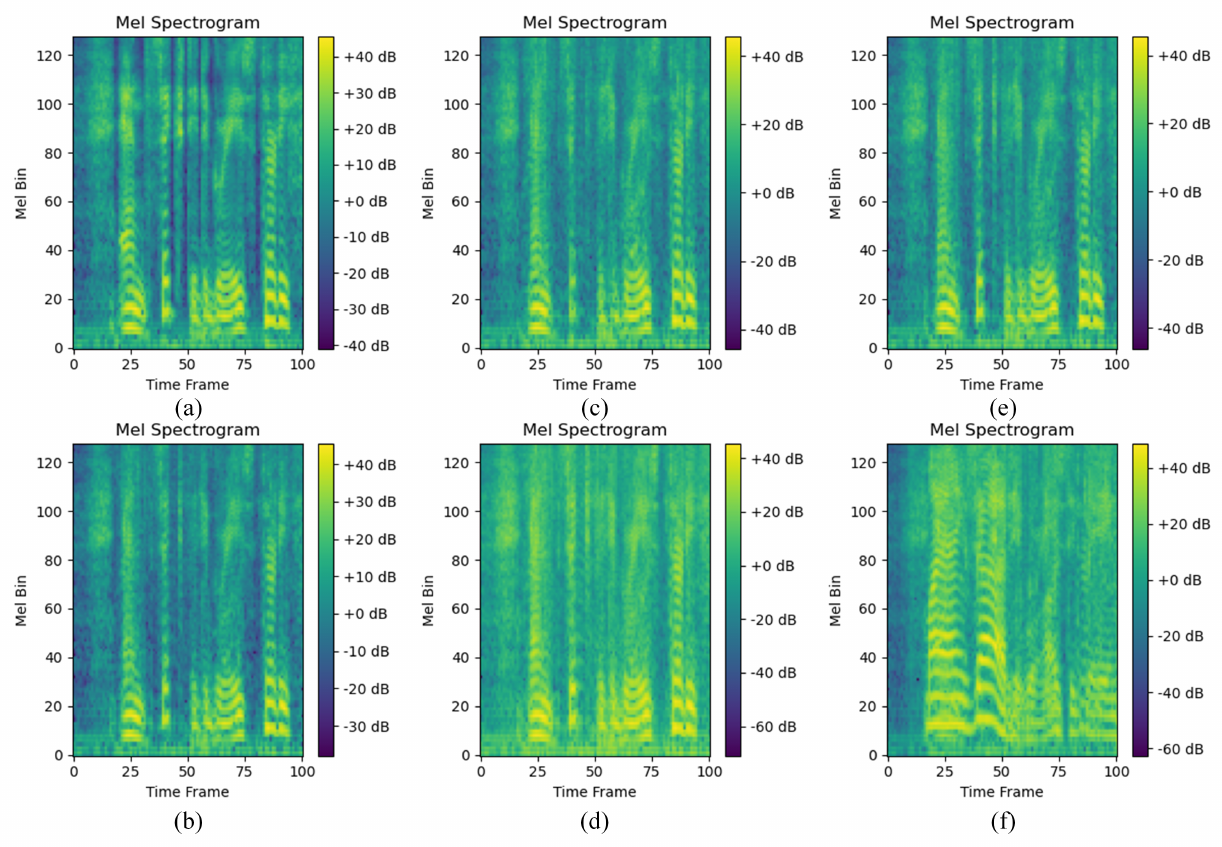}
    \caption{Mel-spectrograms of audio generated from different ablation studies, illustrating the impact of various model components on audio quality.(a) Clean audio, (b) BM-TSE (Full), (c) BM-TSE (w/o ASG), (d) BM-TSE (w/o $\mathcal{L}_{SID}$), (e) BM-TSE (w/o SConv), (f) BM-TSE (w/o LS-TConv)}
    \label{fig:mel}
\end{figure}

As shown in Fig.\ref{fig:mel}, the provided Mel-spectrograms effectively illustrate the impact of an ablation study on the BM-TSE model's audio generation quality. The clean audio in Figure (a) serves as the baseline, showing rich, well-defined spectral details. Our high-fidelity benchmark is the complete BM-TSE model, shown in Figure (b), which demonstrates the full model's effectiveness. The ablation results highlight the critical role of each component. Removing Audio Semantic Guidance (ASG) in Figure (c) leads to a loss of high-frequency detail. Excluding the Speaker Identity Loss ($\mathcal{L}_{SID}$) in Figure (d) introduces noticeable spectral distortion. The absence of Spectral Convolution (SConv) in Figure (e) causes disruptive, blotchy artifacts, while a lack of Long-Short Term Temporal Convolution (LS-TConv) in Figure (f) blurs the temporal dynamics, significantly degrading the audio's coherence. These comparisons demonstrate that each component of the model is essential for generating high-quality, coherent, and artifact-free audio.
\vspace{-1pt}
\section{Conclusion}
\label{sec:conclusion}
In this paper, we introduced BM-TSE, a novel framework designed for personalized and high-fidelity neuro-steered target speaker extraction. BM-TSE integrates a robust ASG-enhanced spatio-temporal encoder to address signal non-stationarity. More centrally, it features a brainprint modulation mechanism where a unified brainmap embedding, jointly supervised by subject identification and attention decoding, actively refines the speech separation process. Extensive experiments on the KUL and Cocktail Party datasets demonstrate that BM-TSE achieves state-of-the-art performance, significantly outperforming existing methods. Our findings validate that leveraging a user's unique neural signature not merely for classification but as a dynamic, modulatory signal within the core audio processing pipeline is a powerful and effective strategy. This approach paves the way for a new generation of truly personalized, neuro-steered assistive technologies.

\vfill\pagebreak

\clearpage


\bibliographystyle{IEEEbib}
\bibliography{strings}

\end{document}